\documentclass{article}
\usepackage{pdfsync}
\usepackage{graphics}
\usepackage{mathptmx}  
\usepackage{float}
\usepackage{graphicx}
\usepackage{natbib}
\usepackage{amsmath}
\usepackage{url}
\usepackage[T1]{fontenc}
\usepackage[utf8]{inputenc}
\usepackage{authblk}
\usepackage{setspace}
\newcommand{\p}{{\rm Pr}}

\begin{document}

\title{Removing batch effects for prediction problems with frozen surrogate variable analysis}
\author[1]{Hilary S. Parker}
\author[2]{H\'{e}ctor Corrada Bravo}
\author[1]{Jeffrey T. Leek\thanks{jleek@jhsph.edu}}
\affil[1]{Department of Biostatistics, Johns Hopkins Bloomberg School of Public Health, Baltimore, MD, 21205}
\affil[2]{Center for Bioinformatics and Computational Biology, Department of Computer Science, University of Maryland, College Park, Maryland, 20742}

\renewcommand\Authands{ and }

  \maketitle

\begin{abstract}
{Batch effects are responsible for the failure of promising genomic prognostic signatures, major ambiguities in published genomic results, and retractions of widely-publicized findings. Batch effect corrections have been developed to remove these artifacts, but they are designed to be used in population studies. But genomic technologies are beginning to be used in clinical applications where samples are analyzed one at a time for diagnostic, prognostic, and predictive applications. There are currently no batch correction methods that have been developed specifically for prediction. In this paper, we propose an new method called frozen surrogate variable analysis (fSVA) that borrows strength from a training set for individual sample batch correction. We show that fSVA improves prediction accuracy in simulations and in public genomic studies.  fSVA is available as part of the \texttt{sva} {\it Bioconductor} package.}
{genomics; prediction; batch effects; personalized medicine; surrogate variable analysis}
\end{abstract}

\doublespacing

\section{Introduction}
Genomic technologies were originally developed and applied for basic science research and hypothesis generation \cite{Eisen1998}. As these technologies mature, they are increasingly being used as clinical tools for diagnosis or prognosis \cite{Chan2011}. The high-dimensional measurements made by microarrays can be used to classify patients into predictive, prognostic, or diagnostic groups. Despite the incredible clinical promise of these technologies there have only been a few signatures that have successfully been translated into the clinic.

One of the reasons for the relatively low rate of success is the impact of unmeasured technological or biological confounders. These artifacts are collectively referred to as ``batch effects'' because the processing date, or batch, is the most commonly measured surrogate for these unmeasured variables in genomic studies \cite{Scharpf2011,Johnson2007b,Walker2008}. The umbrella term batch effects also refers to any unmeasured variables that can vary from experiment to experiment, ranging from the technician who performs the experiment to the temperature and ozone levels that day \cite{Lander1999,Fare2003}. 

Batch effects are responsible for the failure of promising genomic prognostic signatures \cite{Baggerly2004a,Baggerly2005}, major ambiguities in published genomic results \cite{Spielman2007,Akey2007}, and retractions of widely-publicized findings \cite{Sebastiani2010,Lambert2012}. In many experiments, the signal from these unmeasured confounders is larger than the biological signal of interest \cite{Leek2010}. But the impact of batch effects on prediction problems has only recently been demonstrated \cite{Parker2012,Luo2010}. Batch effects were also recognized as a significant hurdle in the development of personalized genomic biomarkers in the Institute of Medicine's report on clinical genomics \cite{Micheel2012}. 

While a number of methods have been developed for removing batch effects in population-based genomic studies \cite{Johnson2007b,Gagnon-Bartsch2011,Leek2007,Leek2010,Walker2008}, there is currently no method for removing batch effects for prediction problems. There are two key differences between population level corrections and corrections designed for prediction problems. First, population level corrections assume that the biological groups of interest are known in advance. In prediction problems, the goal is to predict the biological group. Second, in prediction problems, new samples are observed one at a time, so the surrogate batch variable will have a unique and unknown value for each sample. 

Here we propose frozen Surrogate Variable Analysis (fSVA) as a method for batch correction in prediction problems. fSVA borrows strength from a reference database to address the challenges unique to batch correction for prediction. The fSVA approach has two main components. First, surrogate variable analysis (SVA) is used to correct for batch effects in the training database. Any standard classification algorithm can then be applied to build a classifier based on this clean training data set. Second, probability weights and coefficients estimated on the training database are used to remove batch effects in new samples. The classifier trained on the clean database can then be applied to these cleaned samples for prediction. 

We show with simulated data that the fSVA approach leads to substantial improvement in predictive accuracy when unmeasured variables are correlated with biological outcomes. We also apply fSVA to multiple publicly available microarray data sets and show improvements in prediction accuracy after correcting for batch. The methods developed in this paper have been implemented in the freely available \texttt{sva} Bioconductor package.

\section{Frozen Surrogate Variable Analysis methodology}

\subsection{Removing batch effects from the training set}
The first step in batch correction for prediction problems is to remove batch effects from the training set. In the training set, the biological groups are known. This setting is similar to the population genomics setting and we can use a model for gene expression data originally developed for population correction of unmeasured confounders. If there are $m$ measured features and $n$ samples in the training set, we let $X_{m \times n}$ be the matrix of feature data, where $x_{ij}$ is the value of feature $i$ for sample $j$. For convenience, we will refer to $X$ as the expression matrix for the remainder of the paper. However, our methods can be generally applied to any set of features, including measures of protein abundance, gene expression, or DNA methylation. 

We propose a linear model for the relationship between the expression levels and the outcome of interest $y_j$: $x_{ij} = b_0 + \sum_{k=1}^{p_1} s_k(y_j) + e_{ij}$. The $s_k(\cdot)$ are a set of basis functions parameterizing the relationship between expression and outcome. If the prediction problem is two-class, then $p_1 =1$ and $s_1(y_j) = 1(y_j = 1)$ is an indicator function that sample $j$ belongs to class one. In a multi-class prediction problem $k > 1$ and the $s_k(\cdot)$ may represent a factor model for each class. In matrix form this model can be written \cite{Leek2007,Leek2008}. 
\begin{equation}
X = BS + E
\label{eqn_before_batch}
\end{equation}
where $S_{p_1 \times n}$ is a model matrix of $p_1$ biological variables of interest for the $n$ samples, $B_{m \times p_1}$ are the coefficients for these variables, and  $E_{m \times n}$ is the matrix of errors. 

In genomic studies, the error term $E$ is not independent across samples \cite{Johnson2007b,Leek2007,Leek2008,Walker2008,Friguet2009,Leek2010,Gagnon-Bartsch2011}.  That is, there is still correlation between rows of $E$ after accounting for the model $S$. The correlation is due to unmeasured and unwanted factors such as batch. We can modify model \ref{eqn_before_batch} to account for these measured biological factors and unmeasured biological and non-biological factors:

\begin{equation}
X = BS + \Gamma G + U
\label{full_expression_matrix}
\end{equation}

where $G_{p_2 \times n}$ is a $p_2 \times n$ random matrix, called a dependence kernel \cite{Leek2008} that parameterizes the effect of unmeasured confounders, $\Gamma_{m \times p_2}$ is the $m \times p_2$ matrix of coefficients for $G$, and $U_{m\times n}$ is the $m \times n$ matrix of independent measurement errors. We previously demonstrated that such a decomposition of the variance exists under general conditions typically satisfied in population genomic experiments \cite{Leek2008}. 

In the training set, the biological classes are known, so $S$ is known and fixed. But the matrices $B$, $\Gamma$ and $G$ must be estimated. fSVA first performs surrogate variable analysis (SVA) on the training database in order to identify surrogates for batch effects in the training samples. The training set can be ``cleaned'' of batch effects by regressing the effect of the surrogate variables out of the data for each feature. Any classification algorithm can then be developed on the basis of the clean training data set. 

SVA is an iterative algorithm that alternates between two steps. First SVA estimates the probabilities $\pi_{i\gamma} = \p(\gamma_{i\cdot} \neq \vec{0}| X, S, \hat{G}), \pi_{ib} = \p(b_{i\cdot} \neq \vec{0}| \gamma_{i\cdot} \neq 0, X, S, \hat{G})$ using an empirical Bayes' estimation procedure \cite{Leek2008,Efron2004b,Storey2005}. These probabilities are then combined to define an estimate of the probability that a gene is associated with unmeasured confounders, but not with the group outcome
\begin{eqnarray*}
\pi_{iw} &=& \p(b_{i\cdot} = \vec{0} \; \& \; \gamma_{i\cdot} \;\neq\; \vec{0} | X,S,\hat{G})\\
& = & \p(b_{i\cdot} \; = \; \vec{0}| \gamma_{i\cdot} \; \neq\; 0, X, S, \hat{G}) \p(\gamma_{i\cdot} \; \neq\; \vec{0}| X, S, \hat{G})\\
&=& (1-\pi_{ib})\pi_{i\gamma}
\end{eqnarray*}
The second step of the SVA algorithm weighs each row of the expression matrix $X$ by the corresponding probability weight $\hat{pi}_{iw}$ and performs a singular value decomposition of the weighted matrix. Letting $\hat{w}_{ii} = \hat{\pi}_{iw}$ the decomposition can be written $\hat{W}X = UDV^T$. After iterating between these two steps, the first $p_2$ weighted left singular vectors of $X$ are used as estimates of $G$. An estimate of $p_2$ can be obtained either through permutation \cite{Buja1992} or asymptotic \cite{Leek2011} approaches. 

Once estimates $\hat{G}$ have been obtained, it is possible to fit the regression model in equation \ref{full_expression_matrix} using standard least squares. The result are estimates for the coefficients $\hat{B}$ and $\hat{\Gamma}$. Batch effects can be removed from the training set by setting $\hat{X}^{clean} = X - \hat{\Gamma}\hat{G}$. Any standard prediction algorithm can then be applied to $\hat{X}^{clean}$ to develop a classifier based on batch-free genomic data. The result is a prediction function $f(\hat{X}_{\cdot j}^{clean})$ that predicts the outcome variable $y_j$ based on the clean expression matrix. 

\subsection{Removing batch effects from new samples}

Removing batch effects from the training database is accomplished using standard population genomic SVA batch correction. But the application of classifiers to new genomic samples requires batch correction of individual samples when both the batch and outcome variables are unknown. The fSVA algorithm borrows strength from the training database to perform this batch correction. 

To remove batch effects from a new sample $X_{\cdot j'}$, it is first appended to the training data to create an augmented expression matrix $X^{j'} = [X X_{\cdot j'}]$ where $[\cdot]$ denotes concatenation of columns. To estimate the values of $G$ for the new sample, fSVA uses a weighted singular value decomposition, using the probability weights estimated from the training database $\hat{W}X^{j'} = U^{j'}D^{j'}V^{j'T}$. The result is an estimate $\hat{G}^{j'}$ that includes a column for the new sample. 

To remove batch effects from the new sample, fSVA uses the coefficients estimated from the training database $\hat{\Gamma}$ and the estimated surrogate variables: $\hat{X}^{clean j'} = X^{j'} - \hat{\Gamma} \hat{G}^{j'}$. If there are $n$ training samples, then the $(n+1)$st column of $X^{clean j'}$ represents the new clean sample. The classifier built on the clean training data can be applied to this clean data set to classify the new sample.

\section{Fast fSVA methodology}

fSVA requires that a new singular value decomposition be applied to the augmented expression matrix once for each new sample. Although this is somewhat computationally intensive, in typical personalized medicine applications, sample collection and processing will be spread over a long period of time. In this setting, computational time is not of critical concern. However, for evaluating the fSVA methodology or developing new classifiers using cross-validation, it is important to be able to quickly calculate clean expression values for test samples. 

We propose an approximate fSVA algorithm that greatly reduces computing time by performing a streaming singular value decomposition \cite{Warmuth2007,Warmuth2008}.  The basic idea behind our computation speed-up is to perform the singular value decomposition once on the training data, save the left singular vectors and singular values, and use them to calculate approximate values for the right singular values in new samples. 

When removing batch effects from the training data, the last step is a weighted singular value decomposition of the training expression matrix $WX = UDV^T$. After convergence, the first $p_2$ columns of the matrix $V$ are the surrogate variables for the training set. Since $U$ and $V$ are orthonormal matrices, we can write $V^T = D^{-1} U^T W X$. The matrix $P = D^{-1}U^TW$ projects the columns of $X$ onto the right singular vectors $V^T$.  Pre-multiplying a set of new samples $X^{new}$ by $P$ results in an estimate of the singular values for the new samples: $\hat{V}^{Tnew} = P^TX^{new}$. The surrogate variable estimates for the new samples consist of the first $p_2$ columns of $\hat{V}^{Tnew}$. We obtain clean data for the new samples using the estimated coefficients from the training set, identical to the calculation for the exact fSVA algorithm: $\hat{X}^{clean,new} = X^{new} - \hat{\Gamma} \hat{G}^{new}$. 

Estimates obtained using this approximate algorithm are not identical to those obtained using the exact fSVA algorithm. The projection matrix used in the approximation, $P^T$, is calculated using only the samples in the training set. However, there is only a one-sample difference between the projection calculated in the training set and the projection that would be obtained with exact fSVA. As the training set size grows, the approximation is closer and closer to the answer that would be obtained from the exact algorithm. For smaller databases, there is less computational burden in calculating the exact estimates. However, for large training sets, the computational savings can be dramatic, as described in the simulation below.

\section{Simulation Results}
We performed a simulation examining the benefit of fSVA in prediction problems.  In order to do this, we simulated data using equation \ref{full_expression_matrix} under different distributions of each parameter. We also discretized the probability weights $\pi_{i\gamma}$ and $\pi_{ib}$, and varied the distribution of these probability weights (Table \ref{dist_table}).  We crafted these simulations to mimic scenarios with a subtle outcome and a strong batch effect, which is frequently the case in genomic data.

\begin{table}[ht]
\begin{center}
\begin{tabular}{c|cc}
	& Parameter Distributions & Affected Features \\
\hline
 			& $B \sim N(0,1)$		& $50\%$ batch-affected \\
Scenario 1 	& $\Gamma \sim N(0,3)$	& $50\%$ outcome-affected \\
 			& $U \sim N(0,2)$		& $40\%$ affected by batch and outcome \\
\hline 
			& $B \sim N(0,1)$		& $50\%$ batch-affected \\
Scenario 2 	& $\Gamma \sim N(0,4)$	& $50\%$ outcome-affected \\
 			& $U \sim N(0,3)$		& $40\%$ affected by batch and outcome \\
\hline
			& $B \sim N(0,1)$		& $80\%$ batch-affected \\
Scenario 3 	& $\Gamma \sim N(0,4)$	& $80\%$ outcome-affected \\
 			& $U \sim N(0,3)$		& $50\%$ affected by batch and outcome \\
\end{tabular}
\caption{\textbf{Specifications for the three simulation scenarios used to show the performance of fSVA.} We performed three simulations under slightly different parameterizations to show the effectiveness of fSVA in improving prediction accuracy.  Parameters from equation \ref{full_expression_matrix} were simulated using the distributions specified in this table. Additionally, the percentage of features in the simulation affected by batch, outcome, or both are as indicated in this table. Results from these simulations can be found in Figure \ref{simres}.}
\end{center}
\label{dist_table}
\end{table}

We also specified that each data set have two batches and two outcomes, each equally distributed.  We varied the amount of confounding between batch and outcome in the simulated database, from a Pearson's correlation of $0$ to a correlation of over $0.90$.  For the simulated new samples, the batch and outcome were always uncorrelated.

We simulated 100 database samples and 100 new samples using the parameters described above. Each sample had $10000$ features. As a control, for each iteration in addition to performing fSVA correction, we performed SVA correction on the simulated database alone, and also performed prediction with no batch correction on the simulated database or new samples.

To quantify the effect that fSVA had on prediction, we performed exact fSVA as described above on the simulated database and new samples.  We then performed Prediction Analysis of Microarrays (PAM), a commonly used method for classifying microarrays \cite{Tibshirani2002}.  The PAM prediction model was built on the SVA-corrected database, and then used to predict the outcomes on the fSVA-corrected new samples.  Each simulation was repeated 100 times for robustness.

\begin{figure}[H]
\centering
\includegraphics[width=4.5in]{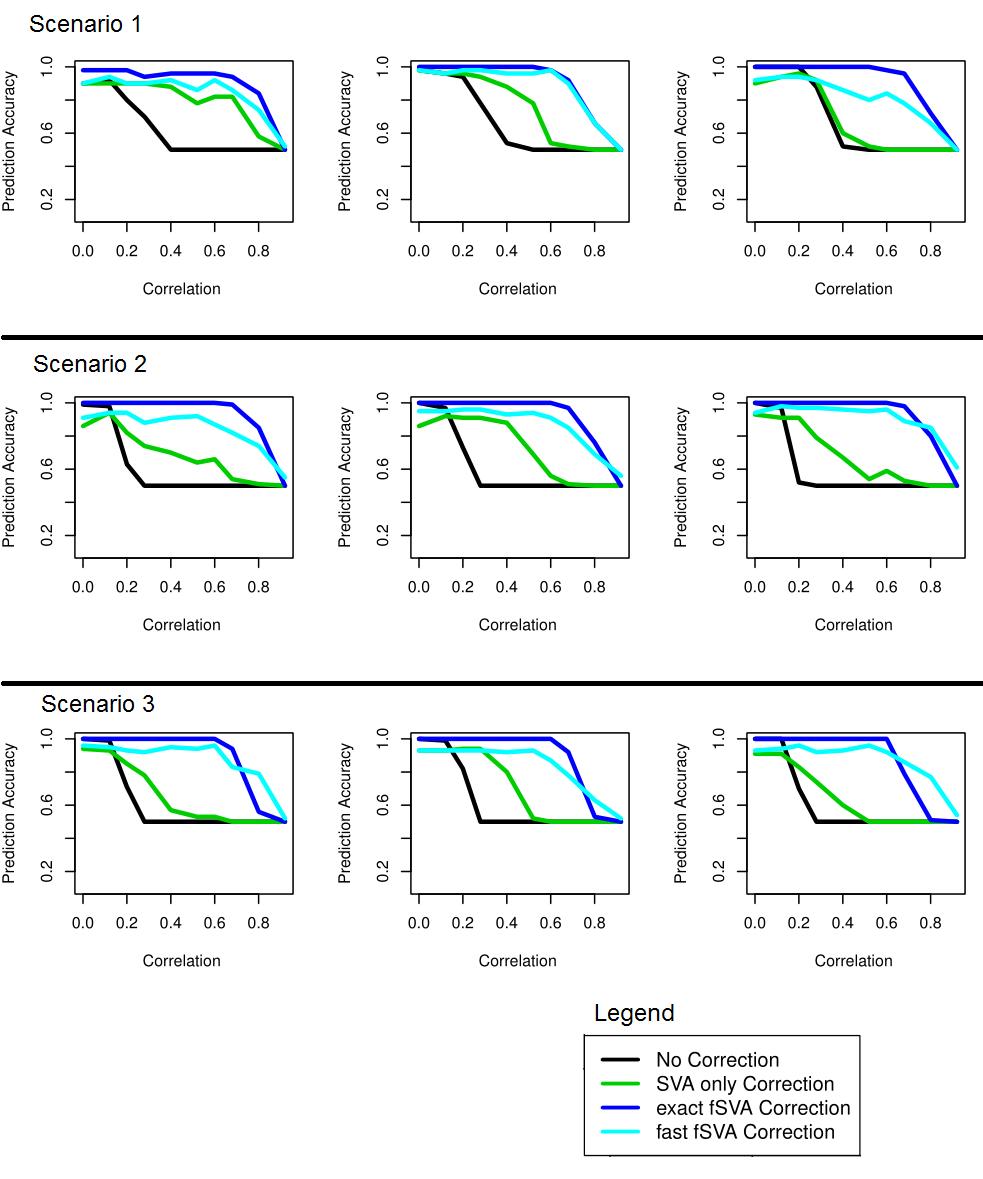}
\caption{\textbf{fSVA improves prediction accuracy of simulated datasets.} We created simulated datasets (consisting of a database and new samples) using model \ref{full_expression_matrix} and tested the prediction accuracy of these using \texttt{R}.  For each simulated data set we performed either exact fSVA correction, fast fSVA correction, SVA correction on the database only, or no correction. We performed 100 iterations on each simulation scenario described in table \ref{dist_table}.  Using the \texttt{sample} function in the \texttt{R} package, we randomly chose three iterations from each of the three scenarios to display here.
}
\label{simres}
\end{figure}

We found that in general the prediction accuracy measures for different iterations of a simulation varied highly, but the ordinality remained relatively constant. Therefore to display results we randomly selected three graphs from each of the scenarios, using the \texttt{sample} function in \texttt{R} (Figure \ref{simres}). Each of the graphs from the 100 iteractions for each scenario can be found on the author's website.

We found that fSVA improved the prediction accuracy in all of our simulations (Figure \ref{simres}). Interestingly, exact fSVA generally outperformed fast fSVA at all of the correlation levels except the highest correlation levels. However both fSVA methods out-performed our control of performing SVA on the database alone. Additionally any method of batch-correction generally outperformed no batch correction whatsoever.  

When the batch and outcome were not correlated, we saw ambiguous performance from using fSVA.  This is not unexpected since it has been shown that in scenarios with no confounding between batch and outcome, batch has a minimal effect on prediction accuracy \cite{Parker2012}.  When databases had extreme confounding between batch and outcome (above $0.85$) we saw the benefits of all the batch-correction methods drop off.  This is because in these situations, SVA on the database cannot differentiate batch and outcome in the database.

While in each of the simulations there was an accuracy cost to using fast fSVA versus exact fSVA, the computational time savings was dramatic. In the scenario described, with 100 samples in the database and 100 new samples, the wall-clock computational time using a standard desktop computer for exact fSVA was 133.9 seconds, versus just 1.3 seconds for fast fSVA. Using 50 samples in the database and 50 new samples, exact fSVA required 17.9 seconds versus 0.4 seconds for fast fSVA. We encourage users to consider both the accuracy and computational times when selecting which algorithm to use for a particular data set.

\section{Results from Microarray Studies}
We examined the effect that fsva had on several microarray studies, obtained from the Gene Expression Omnibus (GEO) website \cite{Edgar2002}.  All except three of the studies were preprocessed/standardized as described previously.  Three of the studies (GSE2034, GSE2603, GSE2990) were obtained from GEO and fRMA-normalized.

Each of the studies was randomly divided into equally-sized ``database'' and ``new sample'' subsets.  We SVA-corrected the database subset, and then built a predictive model (PAM) on that corrected data.  We then performed fSVA correction on the new samples.  After performing fSVA correction, we measured the prediction accuracy of the model built on the database. This process was iterated 100 times for each study to obtain confidence intervals. This method is virtually identical to the simulation described above. 

Results from this process can be found below (Table \ref{studytab}).  Five of the studies showed significant improvement using fSVA.  One study showed marginal improvement, with its 95\% confidence interval overlapped zero.  Three studies showed a cost to using fSVA, though in all three cases the 95\% confidence interval for the true cost overlapped zero.

\begin{table}[ht]
\begin{center}
\begin{tabular}{lccc}
  \hline
Study Name & No Correction (CI) & Exact fSVA (CI) & Improvement (CI) \\
  \hline
GSE10927 & \textbf{0.97} (0.96, 0.97) & \textbf{0.98} (0.98, 0.99) & \textbf{0.02} (0.01, 0.02) \\
GSE13041 & \textbf{0.61} (0.59, 0.63) & \textbf{0.69} (0.66, 0.71) & \textbf{0.07} (0.05, 0.10) \\
GSE13911 & \textbf{0.93} (0.93, 0.94) & \textbf{0.94} (0.94, 0.95) & \textbf{0.01} (0.00, 0.01) \\
GSE2034 & \textbf{0.51} (0.49, 0.52) & \textbf{0.54} (0.52, 0.56) & \textbf{0.03} (0.01, 0.05) \\
GSE2603 & \textbf{0.68} (0.66, 0.70) & \textbf{0.66} (0.64, 0.67) & \textbf{-0.02} (-0.04, 0.00) \\
GSE2990 & \textbf{0.59} (0.58, 0.61) & \textbf{0.58} (0.56, 0.60) & \textbf{-0.02} (-0.04, 0.00) \\
GSE4183 & \textbf{0.89} (0.88, 0.91) & \textbf{0.88} (0.86, 0.89) & \textbf{-0.02} (-0.03, 0.00) \\
GSE6764 & \textbf{0.74} (0.72, 0.76) & \textbf{0.75} (0.72, 0.77) & \textbf{0.01} (-0.01, 0.04) \\
GSE7696 & \textbf{0.78} (0.76, 0.79) & \textbf{0.80} (0.78, 0.81) & \textbf{0.02} (0.01, 0.04) \\
   \hline
\end{tabular}
\caption{\textbf{fSVA improves prediction accuracy in 5 of the 9 studies examined.} The remaining 4 studies showed indeterminate results since the 95\% confidence intervals overlapped zero.  In order to find the prediction accuracy results, each of the studies was randomly divided into ``database samples'' and ``new samples''.  fSVA-correction was then performed as described above.  We then built a predictive model (PAM) on the database and tested the prediction accuracy on the new samples.}
\end{center}
\label{studytab}
\end{table}

\section{Conclusions}

Batch effects have been recognized as a crucial hurdle for population genomics experiments \cite{Leek2010,Parker2012}. They have also been recognized as a critical hurdle in developing genomic signatures for personalized medicine \cite{Micheel2012}. Here we have introduced the first batch correction method specifically developed for prediction problems. Our approach borrows strength from a training set to infer and remove batch effects in individual clinical samples. 

We have demonstrated the power of our approach in both simulated and real gene expression microarray data. However, our approach depends on similarity between the training set and the test samples, both in terms of the genes affected and the estimated coefficients. In small training sets, these assumptions may be violated. Similarly, training sets that show near perfect correlation between batch variables and biological classes represent an extreme case that can not be directly corrected using fSVA. An interesting avenue for future research is the use of publicly available microarray data to build increasingly large training databases for batch removal. 

The methods we have developed here are available as part of the \texttt{sva} Bioconductor package \cite{Leek2012b}.


\bibliographystyle{plain}

\begin{thebibliography}{30}
\providecommand{\natexlab}[1]{#1}
\providecommand{\url}[1]{\texttt{#1}}
\expandafter\ifx\csname urlstyle\endcsname\relax
  \providecommand{\doi}[1]{doi: #1}\else
  \providecommand{\doi}{doi: \begingroup \urlstyle{rm}\Url}\fi

\bibitem[Akey et~al.(2007)Akey, Biswas, Leek, and Storey]{Akey2007}
Joshua~M Akey, Shameek Biswas, Jeffrey~T Leek, and John~D Storey.
\newblock {On the design and analysis of gene expression studies in human
  populations.}
\newblock \emph{Nature Genetics}, 39\penalty0 (7):\penalty0 807--8; author
  reply 808--9, July 2007.
\newblock ISSN 1061-4036.
\newblock \doi{10.1038/ng0707-807}.

\bibitem[Baggerly et~al.(2004)Baggerly, Morris, and Coombes]{Baggerly2004a}
Keith~a Baggerly, Jeffrey~S Morris, and Kevin~R Coombes.
\newblock {Reproducibility of SELDI-TOF protein patterns in serum: comparing
  datasets from different experiments.}
\newblock \emph{Bioinformatics (Oxford, England)}, 20\penalty0 (5):\penalty0
  777--85, March 2004.
\newblock ISSN 1367-4803.
\newblock \doi{10.1093/bioinformatics/btg484}.

\bibitem[Baggerly et~al.(2005)Baggerly, Coombes, and Morris]{Baggerly2005}
Keith~a Baggerly, Kevin~R Coombes, and Jeffrey~S Morris.
\newblock {Bias, randomization, and ovarian proteomic data: a reply to
  "producers and consumers"}.
\newblock \emph{Cancer informatics}, 1\penalty0 (2003):\penalty0 9--14, January
  2005.
\newblock ISSN 1176-9351.

\bibitem[Buja and Eyuboglu(1992)]{Buja1992}
Andreas Buja and Nermin Eyuboglu.
\newblock {Remarks on Parallel Analysis}.
\newblock \emph{Multivariate Behavioral Research}, 27\penalty0 (4):\penalty0
  509--540, 1992.
\newblock \doi{10.1207/s15327906mbr2704\_2}.

\bibitem[Chan and Ginsburg(2011)]{Chan2011}
Isaac~S Chan and Geoffrey~S Ginsburg.
\newblock {Personalized medicine: progress and promise.}
\newblock \emph{Annual review of genomics and human genetics}, 12:\penalty0
  217--44, September 2011.
\newblock ISSN 1545-293X.
\newblock \doi{10.1146/annurev-genom-082410-101446}.

\bibitem[Edgar et~al.(2002)Edgar, Domrachev, and Lash]{Edgar2002}
Ron Edgar, Michael Domrachev, and Alex~E Lash.
\newblock {Gene Expression Omnibus: NCBI gene expression and hybridization
  array data repository}.
\newblock \emph{Nucleic Acids Research}, 30\penalty0 (1):\penalty0 207--10,
  January 2002.
\newblock ISSN 1362-4962.
\newblock \doi{10.1093/nar/30.1.207}.

\bibitem[Efron(2004)]{Efron2004b}
Bradley Efron.
\newblock {Large-Scale Simultaneous Hypothesis Testing}.
\newblock \emph{Journal of the American Statistical Association}, 99\penalty0
  (465):\penalty0 96--104, March 2004.
\newblock ISSN 0162-1459.
\newblock \doi{10.1198/016214504000000089}.

\bibitem[Eisen et~al.(1998)Eisen, Spellman, Brown, and Botstein]{Eisen1998}
M~B Eisen, P~T Spellman, P~O Brown, and D~Botstein.
\newblock {Cluster analysis and display of genome-wide expression patterns}.
\newblock \emph{Proceedings of the National Academy of Sciences}, 95\penalty0
  (25):\penalty0 14863--8, December 1998.
\newblock ISSN 0027-8424.
\newblock \doi{10.1073/pnas.95.25.14863}.

\bibitem[Fare et~al.(2003)Fare, Coffey, Dai, He, Kessler, Kilian, Koch,
  LeProust, Marton, Meyer, Stoughton, Tokiwa, and Wang]{Fare2003}
Thomas~L Fare, Ernest~M Coffey, Hongyue Dai, Yudong~D He, Deborah~a Kessler,
  Kristopher~a Kilian, John~E Koch, Eric LeProust, Matthew~J Marton, Michael~R
  Meyer, Roland~B Stoughton, George~Y Tokiwa, and Yanqun Wang.
\newblock {Effects of atmospheric ozone on microarray data quality.}
\newblock \emph{Analytical chemistry}, 75\penalty0 (17):\penalty0 4672--5,
  September 2003.
\newblock ISSN 0003-2700.

\bibitem[Friguet et~al.(2009)Friguet, Kloareg, and Causeur]{Friguet2009}
Chlo\'{e} Friguet, Maela Kloareg, and David Causeur.
\newblock {A Factor Model Approach to Multiple Testing Under Dependence}.
\newblock \emph{Journal of the American Statistical Association}, 104\penalty0
  (488):\penalty0 1406--1415, December 2009.
\newblock ISSN 0162-1459.
\newblock \doi{10.1198/jasa.2009.tm08332}.

\bibitem[Gagnon-Bartsch and Speed(2011)]{Gagnon-Bartsch2011}
Johann~A Gagnon-Bartsch and Terence~P Speed.
\newblock {Using control genes to correct for unwanted variation in microarray
  data}.
\newblock \emph{Biostatistics}, November 2011.
\newblock ISSN 1465-4644.
\newblock \doi{10.1093/biostatistics/kxr034}.

\bibitem[Johnson et~al.(2007)Johnson, Li, and Rabinovic]{Johnson2007b}
W~Evan Johnson, Cheng Li, and Ariel Rabinovic.
\newblock {Adjusting batch effects in microarray expression data using
  empirical Bayes methods.}
\newblock \emph{Biostatistics}, 8\penalty0 (1):\penalty0 118--27, January 2007.
\newblock ISSN 1465-4644.
\newblock \doi{10.1093/biostatistics/kxj037}.

\bibitem[Lambert and Black(2012)]{Lambert2012}
Christophe~G Lambert and Laura~J Black.
\newblock {Learning from our GWAS mistakes: from experimental design to
  scientific method.}
\newblock \emph{Biostatistics (Oxford, England)}, 13\penalty0 (2):\penalty0
  195--203, April 2012.
\newblock ISSN 1468-4357.
\newblock \doi{10.1093/biostatistics/kxr055}.

\bibitem[Lander(1999)]{Lander1999}
E~S Lander.
\newblock {Array of hope.}
\newblock \emph{Nature genetics}, 21\penalty0 (1 Suppl):\penalty0 3--4, January
  1999.
\newblock ISSN 1061-4036.
\newblock \doi{10.1038/4427}.

\bibitem[Leek(2011)]{Leek2011}
Jeffrey~T Leek.
\newblock {Asymptotic conditional singular value decomposition for
  high-dimensional genomic data.}
\newblock \emph{Biometrics}, 67\penalty0 (2):\penalty0 344--52, June 2011.
\newblock ISSN 1541-0420.
\newblock \doi{10.1111/j.1541-0420.2010.01455.x}.

\bibitem[Leek and Storey(2007)]{Leek2007}
Jeffrey~T Leek and John~D Storey.
\newblock {Capturing heterogeneity in gene expression studies by surrogate
  variable analysis.}
\newblock \emph{PLoS Genetics}, 3\penalty0 (9):\penalty0 1724--35, September
  2007.
\newblock ISSN 1553-7404.
\newblock \doi{10.1371/journal.pgen.0030161}.

\bibitem[Leek and Storey(2008)]{Leek2008}
Jeffrey~T Leek and John~D Storey.
\newblock {A general framework for multiple testing dependence}.
\newblock \emph{Proceedings of the National Academy of Sciences of the United
  States of America}, 105\penalty0 (48):\penalty0 18718--23, December 2008.
\newblock ISSN 1091-6490.
\newblock \doi{10.1073/pnas.0808709105}.

\bibitem[Leek et~al.(2010)Leek, Scharpf, Bravo, Simcha, Langmead, Johnson,
  Geman, Baggerly, and Irizarry]{Leek2010}
Jeffrey~T Leek, Robert~B Scharpf, H\'{e}ctor~Corrada Bravo, David Simcha,
  Benjamin Langmead, W~Evan Johnson, Donald Geman, Keith~A Baggerly, and
  Rafael~A Irizarry.
\newblock {Tackling the widespread and critical impact of batch effects in
  high-throughput data.}
\newblock \emph{Nature Reviews Genetics}, 11\penalty0 (10):\penalty0 733--9,
  October 2010.
\newblock ISSN 1471-0064.
\newblock \doi{10.1038/nrg2825}.

\bibitem[Leek et~al.(2012)Leek, Johnson, Parker, Jaffe, and Storey]{Leek2012b}
Jeffrey~T Leek, W~Evan Johnson, Hilary~S Parker, Andrew~E Jaffe, and John~D
  Storey.
\newblock {The sva package for removing batch effects and other unwanted
  variation in high-throughput experiments}.
\newblock \emph{Bioinformatics}, January 2012.
\newblock ISSN 1367-4811.
\newblock \doi{10.1093/bioinformatics/bts034}.

\bibitem[Luo et~al.(2010)Luo, Schumacher, Scherer, Sanoudou, Megherbi, Davison,
  Shi, Tong, Shi, Hong, Zhao, Elloumi, Shi, Thomas, Lin, Tillinghast, Liu,
  Zhou, Herman, Li, Deng, Fang, Bushel, Woods, and Zhang]{Luo2010}
J~Luo, M~Schumacher, A~Scherer, D~Sanoudou, D~Megherbi, T~Davison, T~Shi,
  W~Tong, L~Shi, H~Hong, C~Zhao, F~Elloumi, W~Shi, R~Thomas, S~Lin,
  G~Tillinghast, G~Liu, Y~Zhou, D~Herman, Y~Li, Y~Deng, H~Fang, P~Bushel,
  M~Woods, and J~Zhang.
\newblock {A comparison of batch effect removal methods for enhancement of
  prediction performance using MAQC-II microarray gene expression data}.
\newblock \emph{The Pharmacogenomics Journal}, 10\penalty0 (4):\penalty0
  278--91, August 2010.
\newblock ISSN 1473-1150.
\newblock \doi{10.1038/tpj.2010.57}.

\bibitem[Micheel et~al.(2012)Micheel, Nass, and Omenn]{Micheel2012}
Christine~M Micheel, Sharyl~J Nass, and Gilbert~S Omenn.
\newblock \emph{{Evolution of Translational Omics: Lessons Learned and the Path
  Forward}}.
\newblock The National Academies Press, 2012.
\newblock ISBN 9780309224185.

\bibitem[Parker and Leek(2012)]{Parker2012}
Hilary Parker and Jeffrey~T Leek.
\newblock {The practical effect of batch on genomic prediction The practical
  effect of batch on genomic prediction}.
\newblock \emph{Statistical Applications in Genetics and Molecular Biology},
  2012.
\newblock \doi{10.1515/1544-6115.1766}.

\bibitem[Scharpf et~al.(2011)Scharpf, Ruczinski, Carvalho, Doan, Chakravarti,
  and Irizarry]{Scharpf2011}
Robert~B Scharpf, Ingo Ruczinski, Benilton Carvalho, Betty Doan, Aravinda
  Chakravarti, and Rafael~a Irizarry.
\newblock {A multilevel model to address batch effects in copy number
  estimation using SNP arrays.}
\newblock \emph{Biostatistics (Oxford, England)}, 12\penalty0 (1):\penalty0
  33--50, January 2011.
\newblock ISSN 1468-4357.
\newblock \doi{10.1093/biostatistics/kxq043}.

\bibitem[Sebastiani et~al.(2010)Sebastiani, Solovieff, Puca, Hartley, Melista,
  Dworkis, Wilk, Myers, Steinberg, Baldwin, and Perls]{Sebastiani2010}
Paola Sebastiani, Nadia Solovieff, Annibale Puca, Stephen~W Hartley, Efthymia
  Melista, Daniel~A Dworkis, Jemma~B Wilk, Richard~H Myers, Martin~H Steinberg,
  Clinton~T Baldwin, and Thomas~T Perls.
\newblock {Genetic Signatures of Exceptional Longevity in Humans}.
\newblock \emph{Science}, 2010, 2010.
\newblock \doi{10.1126/science.1190532}.

\bibitem[Spielman et~al.(2007)Spielman, Bastone, Burdick, Morley, Ewens, and
  Cheung]{Spielman2007}
Richard~S Spielman, Laurel~a Bastone, Joshua~T Burdick, Michael Morley,
  Warren~J Ewens, and Vivian~G Cheung.
\newblock {Common genetic variants account for differences in gene expression
  among ethnic groups.}
\newblock \emph{Nature genetics}, 39\penalty0 (2):\penalty0 226--31, February
  2007.
\newblock ISSN 1061-4036.
\newblock \doi{10.1038/ng1955}.

\bibitem[Storey et~al.(2005)Storey, Akey, and Kruglyak]{Storey2005}
John~D Storey, Joshua~M Akey, and Leonid Kruglyak.
\newblock {Multiple locus linkage analysis of genomewide expression in yeast.}
\newblock \emph{PLoS biology}, 3\penalty0 (8):\penalty0 e267, August 2005.
\newblock ISSN 1545-7885.
\newblock \doi{10.1371/journal.pbio.0030267}.

\bibitem[Tibshirani et~al.(2002)Tibshirani, Hastie, Narasimhan, and
  Chu]{Tibshirani2002}
Robert Tibshirani, Trevor Hastie, Balasubramanian Narasimhan, and Gilbert Chu.
\newblock {Diagnosis of multiple cancer types by shrunken centroids of gene
  expression}.
\newblock \emph{Proceedings of the National Academy of Sciences}, 99\penalty0
  (10):\penalty0 6567--72, May 2002.
\newblock ISSN 0027-8424.
\newblock \doi{10.1073/pnas.082099299}.

\bibitem[Walker et~al.(2008)Walker, Liao, Gilbert, Wong, Pollard, McCulloch,
  Lit, and Sharp]{Walker2008}
Wynn~L Walker, Isaac~H Liao, Donald~L Gilbert, Brenda Wong, Katherine~S
  Pollard, Charles~E McCulloch, Lisa Lit, and Frank~R Sharp.
\newblock {Empirical Bayes accomodation of batch-effects in microarray data
  using identical replicate reference samples: application to RNA expression
  profiling of blood from Duchenne muscular dystrophy patients.}
\newblock \emph{BMC genomics}, 9:\penalty0 494, January 2008.
\newblock ISSN 1471-2164.
\newblock \doi{10.1186/1471-2164-9-494}.

\bibitem[Warmuth and Kuzmin(2007)]{Warmuth2007}
Manfred~K Warmuth and Dima Kuzmin.
\newblock {Randomized PCA Algorithms with Regret Bounds that are Logarithmic in
  the Dimension}.
\newblock \emph{Advances in Neural Information Processing Systems},
  19:\penalty0 1481, 2007.

\bibitem[Warmuth and Kuzmin(2008)]{Warmuth2008}
Manfred~K Warmuth and Dima Kuzmin.
\newblock {Randomized Online PCA Algorithms with Regret Bounds that are
  Logarithmic in the Dimension}.
\newblock 9:\penalty0 2287--2320, 2008.
\newblock \doi{10.1.1.133.8332}.

\end{thebibliography}

\end{document}